# Camphor-Engine-Driven Micro-Boat Guides Evolution of Chemical Gardens[1]

Mark Frenkel[1,2], Victor Multanen[2], Roman Grynyov[1], Albina Musin[1], Yelena Bormashenko[2], Edward Bormashenko,[2*]

[1]Ariel University, Natural Sciences Faculty, Physics Department, P.O.B. 3, 407000, Ariel, Israel

[2]Ariel University, Faculty of Engineering, Department of Chemical Engineering, P.O.B. 3, 407000, Ariel, Israel

[*]Corresponding author: Edward Bormashenko E-mail: edward@ariel.ac.il

**ABSTRACT**

A micro-boat self-propelled by a camphor engine, carrying seed crystals of $FeCl_3$, promoted the evolution of chemical gardens when placed on the surface of aqueous solutions of potassium hexacyanoferrate. Inverse chemical gardens (growing from the top downward) were observed. The growth of the "inverse" chemical gardens was slowed down with an increase in the concentration of the potassium hexacyanoferrate. Heliciform precipitates were formed under the self-propulsion of the micro-boat. A phenomenological model, satisfactorily describing the self-locomotion of the camphor-driven micro-boat, is introduced and checked.

## 1   INTRODUCTION

Understanding the autonomous, self-driven motion of micro-swimmers is of a prime importance for the analysis of complex biological, chemical and physical systems.[17-2] For biological systems, such as bacterial colonies, plankton, or fish

---

swarms, swimming droplets can provide a crucial link between simulations and real systems.[1-2]

Self-propulsion of a macroscopic object is achieved by converting external physical energies from the surrounding (energy of the electromagnetic field, chemical or thermal energy and thermal gradients) into mechanical energy.[3-10] Popular mechanisms of self-propulsion are based on exploiting Marangoni thermo- and soluto-capillary flows.[12-22] One the most effective and deeply studied systems based on the Marangoni soluto-capillary driven self-propulsion is a so-called camphor boat.[21, 23-26] The mechanism of the self-locomotion of the camphor boat is related to the flow, inspired by change of the surface tension caused by the dissolution of camphor, followed by its diffusion.[21, 23-26]

In the present paper, we use camphor-based self-propulsion exploited for guided growth of chemical gardens.[27-32] Chemical gardens are multi-scaled, tubular structures arising from a complex, self-organizing non-equilibrium process involving complex chemical and hydrodynamic events. The characteristic spatial scales observed in the study of chemical gardens start from nano-meters[33-34] and extend to centimeters.[26-31] Chemical gardens, described first in 1646 by Johann Glauber, still attract considerable interest among researchers. A broad diversity of systems form chemical gardens, including the chemical processes occurring in Portland cement.[35-36] The generation of electrochemical energy via chemical gardens was discussed in Ref. 37.

The fundamental aspects of chemical gardens' reaction and growth are as important as their applications. Tubular filaments obtained under a chemical garden reaction exhibit curved, helical morphologies, reminiscent of biological forms. The filaments are similar to supposed cyanobacterial microfossils, suggested to be the oldest known terrestrial microfossils.[38] It was also demonstrated that the growth of stalactites and stalagmites in limestone caves are controlled to a great extent by the same phenomena as those occurring in the growth of chemical gardens.[39]

The perplexed hydrodynamics of growing chemical gardens was studied recently in detail by various groups.[40-43] Our group demonstrated recently that chemical gardens may be grown in water polymer solutions, and also under geometrical confinement within microliter droplets.[44-46] Our present paper shows the possibility to grow chemical gardens under the self-propelled motion of a seed crystal driven by a camphor boat.

## 2. EXPERIMENTAL SECTION.

### 2.1. Materials

$FeCl_3$ powder, camphor 96% and potassium hexacyanoferrate (II) trihydrate (($K_4Fe(CN)_6 \cdot 3H_2O$), also known as Yellow Prussiate, were supplied by Sigma–Aldrich. Bi-distilled water (resistivity $\rho \cong 18 M\Omega \cdot cm$) was used for preparing solutions. Aqueous solutions of potassium hexacyanoferrate with the concentrations $c$=0.125-3 wt.% were prepared.

### 2.2. Construction of the camphor-driven, self-propelling micro-boat containing $FeCl_3$ seed crystals.

The self-propelling micro-boat containing $FeCl_3$ seed crystals was prepared with plastic (polyvinylchloride (PVC)) tubing, as shown in Fig. 1a-b (the similar construction of the micro-boat was proposed in Ref. 12). Camphor powder and $FeCl_3$ seed crystals were introduced into opposite ends of the tubing, as shown in Fig. 1a-b. Two boats with the lengths $L_1$= 10 mm and $L_2$= 15 mm were tested. The masses of the PVC tubes $m_{PVC}$ were 0.0325 g ($L_1$) and 0.049 g ($L_2$); the mass of camphor $m_c$ was *ca.* $2.0 \times 10^{-3} g$; the mass of FeCl₃ seed crystals $m_{sc}$ was *ca.* $3.0 \times 10^{-3} g$.

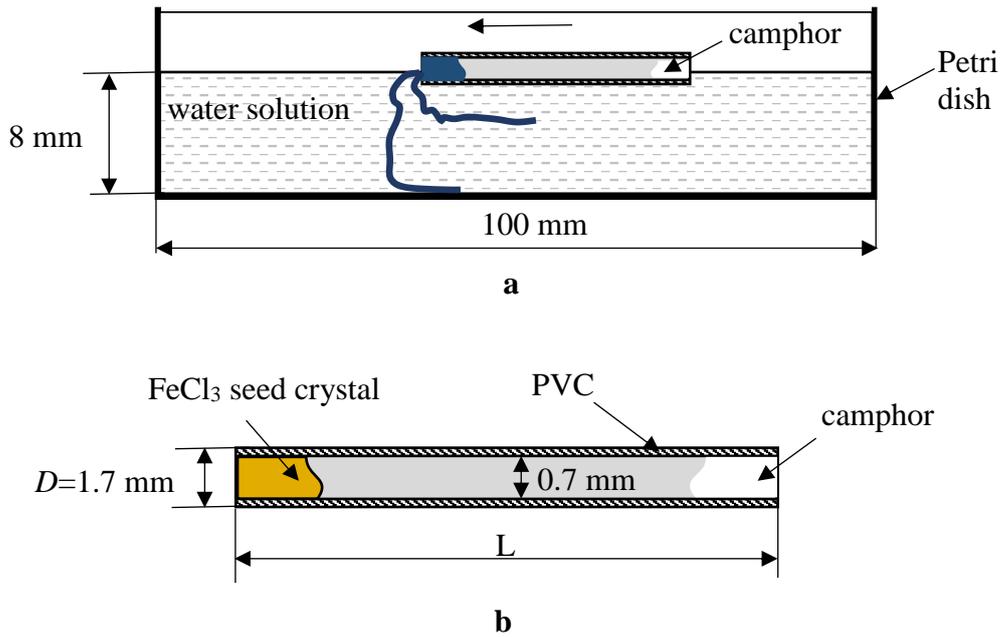

**Figure 1. a.** - A schematic image of the camphor engine driven growth of the chemical gardens. The boat is placed in a Petri dish. Blue wires depict filaments grown under the chemical gardens reaction. **b**. - Construction of the self-propelling micro-boat containing FeCl₃ seed crystals. The PVC tube contains FeCl₃ seed crystals and camphor. The lengths of the tubes $L$ in the experiments were 10mm and 15 mm.

### 2.3. Experimental Methods

The process of self-propelling and the growth of the chemical gardens were registered with the rapid cameras Casio EX-FH20 and the digital microscope BW1008-500X. After capturing the video, the movie was split into separate frames by the VirtualDub software. The frames were treated by a specially developed software, enabling the calculation of the speed of the center mass of the micro-boat.

Surface tensions of potassium hexacyanoferrate solutions with the concentrations varied from 0.125 to 3 wt.% were established using the Ramé-Hart Advanced Goniometer (Model 500-F1) by the pendant droplet method. The dynamic viscosity was measured with an Ostwald-type capillary viscometer consisting of a *U*-shaped glass tube in a controlled temperature bath (25±0.2ºC). The density was established by a Gay-Lussac type Isolab (Germany) pycnometer with the capacity of 10ml. Results of the measurements are summarized in Table 1.

**Table 1**. **The density, surface tension and dynamic viscosity vs. the concentration of $K_4Fe(CN)_6$ water solutions**.

| Concentration, $c$, (wt.%) | Density, $\rho$, ±0.001 $g/cm^3$ | Surface tension, $\gamma$, ±0.05 $mJ/m^2$ | Viscosity, $\eta$, ±0.3×10$^{-4}$ $Pa \times s$ |
|---|---|---|---|
| 0 (pure water) | 0.997 | 71.66 | 8.90 |
| 0.125 | 0.998 | 71.28 | 8.93 |
| 0.25 | 0.999 | 70.80 | 9.00 |
| 0.5 | 1.001 | 71.22 | 9.09 |
| 1 | 1.003 | 71.04 | 9.21 |
| 2 | 1.009 | 70.31 | 9.26 |
| 3 | 1.014 | 70.14 | 9.54 |

The topography of the precipitate was studied with the Carl Zeiss Axiolab A 45 09 09 optical microscope (Germany), and the IDS UI-1490LE-C-HQ digital camera (Germany). Post processing of the stack of images was done with the Zerene Stacker "focus stacking" software. Thermal imaging of the process was carried out with the Therm-App TAS19AQ-1000-HZ thermal camera equipped with a LWIR 6.8 mm f/1.4 lens. Contact angles for the potassium hexacyanoferrate/Petri dish system were established with the Ramé-Hart goniometer (Model 500, USA).

Four series of experiments were carried out, namely:

1) Study of the self-propulsion of the PVC tubing (depicted in Fig. 1A) driven by camphor on distilled water.

2) Study of the self-propulsion of the PVC tubing driven by camphor when placed on potassium hexacyanoferrate aqueous solutions, $c$=0.125-3 wt.% (without $FeCl_3$ seed crystals).

3) Study of the self-propulsion of the PVC tubing containing $FeCl_3$ seed crystals, driven by camphor when the tubing was placed on potassium hexacyanoferrate aqueous solutions, $c$=0.125-3 wt.%. In this case the chemical gardens reaction was observed during the motion of the PVC tubing.

4) Study of the static "inverse" growth of chemical growth, when PVC tubing containing only $FeCl_3$ crystals was placed on potassium hexacyanoferrate aqueous solutions

## 3. RESULTS AND DISCUSSION.

### 3.1. Static growing of chemical gardens.

At first we studied a static (the micro-boat was at rest) growing of chemical gardens with the PVC tube filled at one end with $FeCl_3$ seed crystals, as depicted with Fig. 2. The growth of chemical gardens was observed under various concentrations ($c$=0.125-3 wt.%) of $K_4Fe(CN)_6$ aqueous solutions. The PVC tube containing seed crystals was at rest and the inverse (top-down) growth of chemical gardens was observed for all studied concentrations. Similar downward-growing chemical gardens

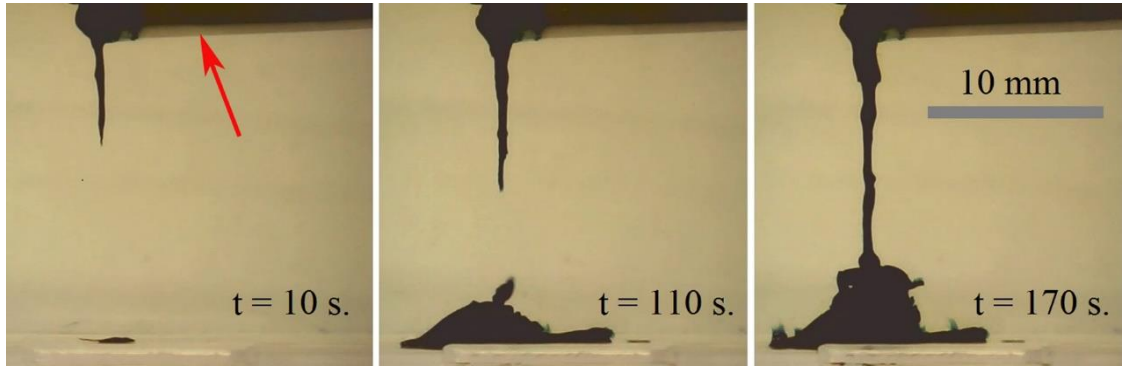

**Figure 2**. Static inverse growth of chemical gardens obtained with PVC tube filled at one end with $FeCl_3$ crystals (the red arrow indicates the location of the floating PVC tube). Concentration of aqueous solutions of potassium hexacyanoferrate $c = 2.5$ wt.%

were already reported in Ref. 44, where the gardens grew in aqueous solutions of polymers.

The optical microscopy images of typical cylindrical filaments constituting the "chemical gardens" are supplied in Supplementary Fig. S1 online. It is noteworthy that the evolution of the gardens was slowed with the increase in the concentration of potassium hexacyanoferrate. The characteristic times of the downward growth of filaments $\tau_1, \tau_2$ were defined as the time span from placing the boat with the seed crystal on the surface of the potassium hexacyanoferrate aqueous solution until the appearance of the filament (as registered with rapid camera and software VirtualDub-1.10.4, used for sequencing of frames), and correspondingly the time measured between its appearance at the end of the tube and time when the filament touched the bottom of the Petri dish. The dependences of $\tau_1, \tau_2$ vs. the concentrations of the solutions are presented in Fig. 3. The data supplied in Table 1, demonstrate that the physical parameters of the aqueous solutions of potassium hexacyanoferrate vary very slightly with the concentration of the solution. Thus, it is reasonable to relate the strong dependences $\tau_1(c), \tau_2(c)$, illustrated in Fig. 3, to the chemical aspects of the evolution of chemical gardens inherent for the studied system, and not to the physical factors.

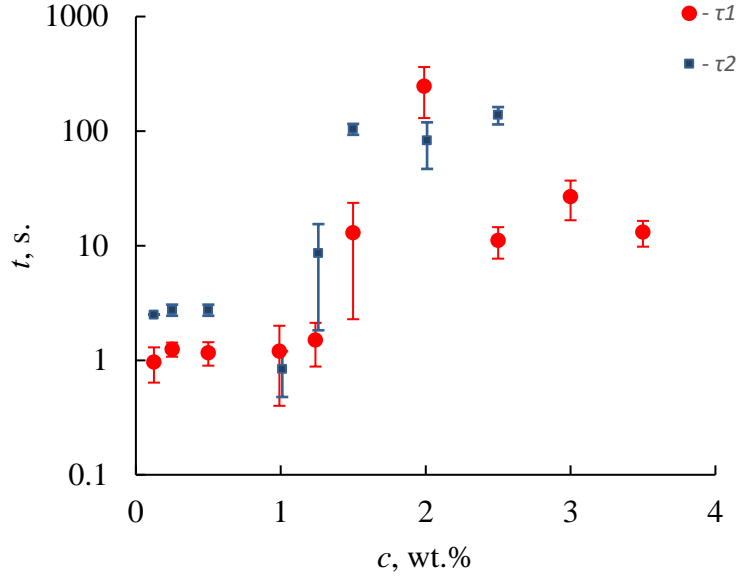

**Figure 3**. Characteristic times of growth of "inverse" chemical gardens $\tau_1$, $\tau_2$ vs. the weight concentration of aqueous solutions of potassium hexacyanoferrate $c$.

It is noteworthy that starting from $c=$ 3 wt.% of the solutions, the filaments growing downward did not touch the bottom of the Petri dish and remained "hanging". A profound understanding of the chemistry of growing "inverse" gardens requires future investigations.

**3.2. The dynamics of growth of chemical gardens inspired by the self-propelled motion of the micro-boat.**

We studied the self-propulsion of the PVC tubing driven by camphor when placed on distilled water and aqueous solutions of potassium hexacyanoferrate; we also investigated the motion of the tubing with and without $FeCl_3$ seed crystals. In all the cases the same temporal pattern of the motion was observed, namely: the short interval of the accelerated self-propulsion was followed by relatively long-term decelerated motion. The dynamics of self-propulsion will be discussed below in detail.

When the micro-boat depicted in Fig. 1a, containing both camphor and seed crystals, was placed on the surface of aqueous solutions of $K_4Fe(CN)_6$, it started a self-propelled motion driven by the soluto-capillary Marangoni flow, owing to the dissolution of the camphor[21-26]. At first the boat moved towards skirting (rim) of the Petri dish and afterwards traveled along the skirting at the distance of *ca*. 1 cm from the rim (see the sequence of images, presented in Fig. 4).

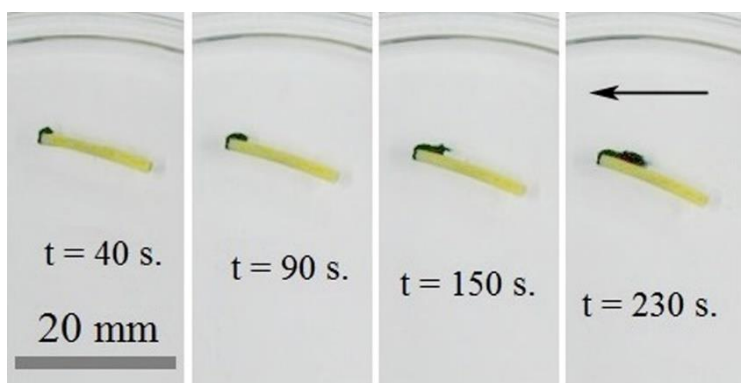

**Figure 4**. Dynamic growing of chemical gardens driven by the self-propelled micro-boat (the black arrow shows the direction of the motion). Concentration of aqueous solutions of potassium hexacyanoferrate $c = 2$ wt.%; $L=15$mm.

In parallel, the chemical garden reaction resulted in the formation of heliciform precipitate, shown with blue in Fig. 5 as taken from a video (see Supplementary Information).

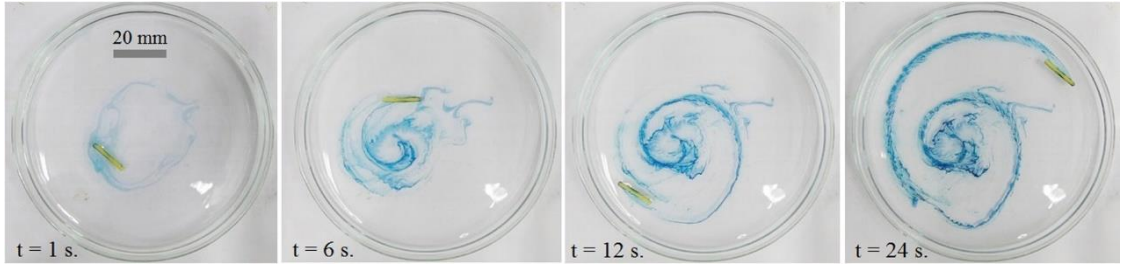

**Figure 5**. Heliciform precipitates formed under the self-propulsion of the micro-boat. $L$=15 mm, $c = 0.25$ wt.%.

The total length and the maximal width of the precipitate (assumed as its featuring geometric characteristics) grew with time. The kinetics of their change is illustrated in Supplementary Fig. S2 online.

Now consider in more detail the dynamics of the self-propulsion, depicted schematically in Fig. 6.

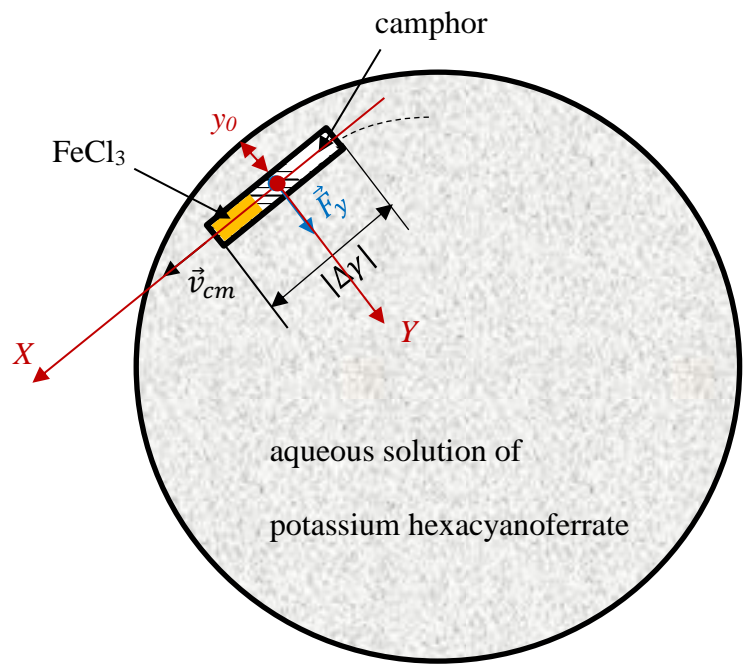

**Figure 6**. Scheme illustrating the self-propulsion. $|\Delta\gamma|$ is the jump in the surface tension driving the self-propulsion.

The camphor-engine-inspired self-propelling of the micro-boat, represented with Fig. 7, accompanied by the growth of chemical gardens, occurs in two distinct stages: accelerated motion which lasts *ca*. 2-10s, followed by decelerated motion which continues *ca*. 40-250 s. The same scales were observed when PVC tubing without seed crystals driven by camphor was placed on aqueous solutions of potassium hexacyanoferrate. However, the time scale of the decelerated motion observed for the self-propulsion of the tubing on the distilled water was *ca*. 1000 s. The change in viscosity of water due to the presence of $K_4Fe(CN)_6$ is negligible (see Table 1). Thus, it is reasonable to suggest that the presence of $K_4Fe(CN)_6$ slows dissolution of camphor, and decreases, in turn, the change of the surface tension, giving rise to the self-propulsion.

The temporal pattern of the motion was the same for micro-boats of various lengths, used in our study, and for various concentrations of aqueous $K_4Fe(CN)_6$ solutions, as shown in Fig. 7. A rigorous analysis of the self-propulsion, accompanied with the chemical gardens reaction is an extremely complicated mathematical task, so we restricted ourselves to a very approximate qualitative treatment of the problem. It is reasonable to relate the twin-stage (accelerated/decelerated) nature of the self-propulsion, illustrated with Fig. 7, to the physical reasoning discussed recently in Refs. 21, 47. Phenomenologically, the self-locomotion of the camphor-driven micro-boat may be explained by eq. (1) (see Refs. 48-49):

$$m\frac{d\vec{v}_{cm}}{dt} = \vec{F}_{fr} + \alpha LD\nabla\gamma = -\chi D\eta\vec{v}_{cm} + \alpha LD\nabla\gamma, \qquad (1)$$

where $m$, $L$, $D$ and $\vec{v}_{cm}$ are the mass, length, diameter and velocity of the center mass of the boat correspondingly, $\gamma$ and $\eta$ are the surface tension and the viscosity of the supporting liquid correspondingly, $\nabla\gamma$ is the gradient of the surface tension due to the dissolution of camphor; $\alpha$ and $\chi$ are the dimensionless, phenomenological coefficients, depending on the shape of self-propelled object (PVC tubing carrying the seed crystal in our case). It should be stressed that the self-propelled micro-boat does not touch the skirting (rim) of the Petri dish; thus, the friction force is estimated as: $\vec{F}_{fr} \cong -\chi D \eta \vec{v}_{cm}$, and we also neglect the contributions to the friction arising from the evolution of the chemical gardens (the influence of the growth of chemical gardens on the time scale of the decelerated motion was minor). The motion occurred under "intermediate Reynolds" numbers $\text{Re} = \frac{v_{cm} D}{\nu}$ (where $\nu$ is the kinematic viscosity of water); indeed, assuming $v_{cm} \cong 3 \times 10^{-2} \frac{m}{s}; D \cong 2 \times 10^{-3} m, \nu \cong 1.0 \times 10^{-6} \frac{m^2}{s}$ yields $\text{Re} \cong 60$.[49-50] In this case, the linear translational and rotational motions of self-propelled boats, are possible, as discussed in Refs. 49-50. Heliciform precipitates formed under the self-propulsion of the micro-boat, shown in Figure 5, hint to co-occurrence of the translational and rotational motions of the boat. When the boat approached the rim of the Petri dish its motion was also guided by the meniscus (i.e the curvature of the liquid/vapor interface), formed in the vicinity of the rim.

The accurate solution of eq. (1) is a challenging task; we assume tentatively that:

$$|\nabla \gamma| \cong \frac{|\Delta \gamma|_0}{L} \exp\left(-\frac{t}{\tau_d}\right), \qquad (2)$$

where $\tau_d$ is the characteristic time of the change in the surface tension due to the dissolution of camphor, followed by its diffusion. When the camphor is placed on the water, the surface tension is decreased rapidly, and then it is slowly decreased with time.[51] $|\Delta \gamma|_0 \cong 5 \, mJ/m^2$ is the initial jump in the surface tension due to the dissolution of the camphor.[52] We do not set into the details of the process of dissolution and diffusion of the camphor, but rather describe the process in a purely phenomenological way with eq. (2). Considering Eq. (2) enables re-shaping of Eq. (1) for the modulus of the velocity of the center of mass of PVC tubing, in the following way:

$$m \frac{d v_{cm}}{dt} = -\chi D \eta v_{cm} + \alpha D |\Delta \gamma|_0 \exp\left(-\frac{t}{\tau_d}\right). \qquad (3)$$

It is convenient to re-shape eq. (3) as given below:

$$\frac{dv_{cm}}{dt} + \frac{1}{\tau_{fr}} v_{cm} = \tilde{a} \exp\left(-\frac{t}{\tau_d}\right) \qquad (4)$$

where $\tau_{fr} \cong m/\chi D\eta$ is the characteristic time of viscous friction-based deceleration of the PVC tubing; $\tilde{a} \cong \alpha D |\Delta\gamma|_0 / m$ is the constant with the dimension of the acceleration. The total mass of the self-propelling system $m$ is comprised of the mass of PVC tubing $m_{PVC}$, the mass of the camphor $m_c$ and the mass of the seed crystals $m_{sc}$ (see the Experimental Section). Thus, the total mass is expressed as $m = m_{PVC} + m_c + m_{sc} = 3.75 - 5.4 \times 10^{-5}$ kg (for PVC boats of various lengths). As it is seen from the quantitative data supplied in the Experimental Section, $m_{PVC} \gg m_c \cong m_{sc} \Rightarrow m \cong m_{PVC}$; hence, for the mass of the self-propelling system we may assume $m \cong \rho_{PVC} L$, where $\rho_{PVC}$ is the length (linear) density of the PVC tubing with the dimensions: $[\rho_{PVC}] = kg/m$ (for the tubes used in our study, $\rho_{PVC} = 3.25 \times 10^{-3} \, kg/m$). This estimation yields for the characteristic time of friction: $\tau_{fr} \cong \rho_{PVC} L / \chi D \eta$, and it is linearly dependent on the length of the PVC tubing $L$.

The true values of the parameters $\alpha, \tilde{a}, \tau_{fr}$ and $\tau_d$ remain unknown, and we consider them as "free" fitting parameters. The solution of the differential equation eq. (4) and considering the initial condition $v_{cm}(t=0) = 0$ yield:

$$v_{cm}(t) = \frac{\tilde{a}}{\tau_{fr}^{-1} - \tau_d^{-1}} [\exp\left(-\frac{t}{\tau_d}\right) - \exp\left(-\frac{t}{\tau_{fr}}\right)], \qquad (5)$$

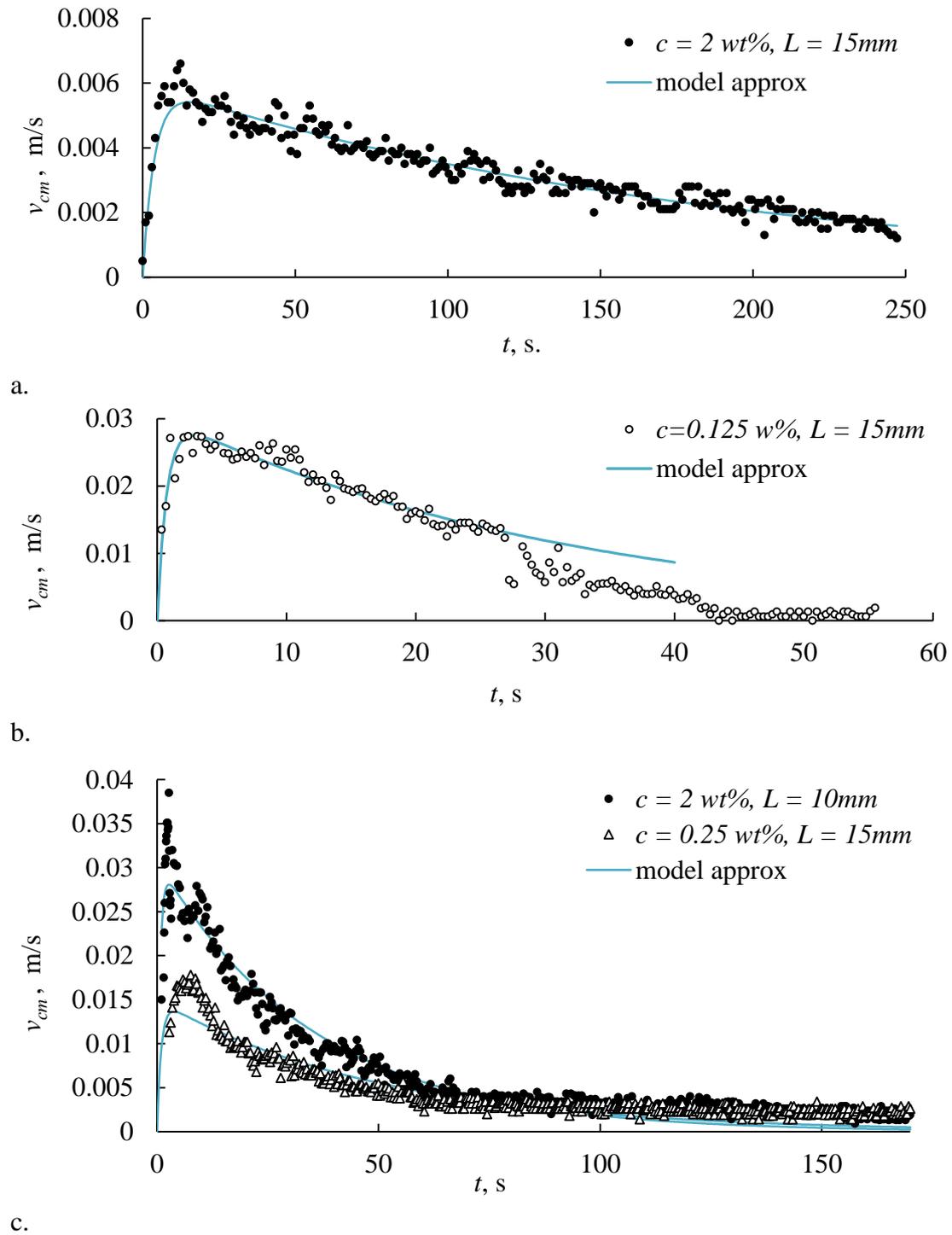

a.

b.

c.

**Figure 7.** The time dependence of the modulus of the velocity of the center of mass of the boat $v_{cm}$ for boats with various lengths moving in water solutions of potassium hexacyanoferrate of various concentrations $c$ wt.%.

Considering $\tilde{a}, \tau_d$ and $\tau_{fr}$ as fitting parameters we fitted the experimental data by the dependence supplied by eq. (5). The resulting fitting curves is shown with a solid, blue lines in Fig. 7. The values of fitting parameters calculated for various experimental conditions are summarized in Supplementary Table S1.

It is recognized from the data supplied in Supplementary Table S1, that the values of the characteristic time scale $\tau_d$ vary within one order of magnitude for various solution concentrations and various lengths of the micro-boat, namely $\tau_d \approx 30-200 s$. It should be emphasized that this time scale is close to the characteristic time of the change in the surface pressure driving the camphor boats, established experimentally in Ref. 21. The characteristic times of viscous dissipation extracted from our experimental data and supplied in Supplementary Table S1 are close, i.e. $\tau_{fr} \approx 1-4s$ for various solution concentrations and lengths of the micro-boat. This result is highly expectable due to the fact that the viscosity of $K_4Fe(CN)_6$ aqueous solutions is only slightly dependent on their concentrations, as is seen from the data, summarized in Table 1. It is also reasonable that the time scale $\tau_{fr}$ grew with the length of the micro-boat as discussed above. The aforementioned arguments justify the phenomenological twin-time model, proposed for the explanation of self-locomotion of the micro-boat.

The parameter $\tilde{a} \cong \alpha D |\Delta\gamma|_0 / m$ appearing in Eq. (4) also is confined within the same order of magnitude for various solution concentrations (see Supplementary Table S1), indicating that the initial jump in the surface tension propelling the boat $|\Delta\gamma|_0$ does not change dramatically with the concentration of $K_4Fe(CN)_6$.

The rotational motion of the micro-boat along the rim of the Petri dish should be addressed within the model developed in Ref. 53 (see Supplementary text and Fig.

S4). It is also noteworthy that when the boat has been initially placed in the vicinity (*ca* 1-2 mm) of the rim of the Petri dish, it started rotational motion along the rim, accompanied with the inverse growth of chemical gardens. Thus, the motion of the boat and consequently growth of chemical gardens may be guided by the shape of the vessel.

One more experimental observation should be discussed. To our best knowledge, the thermal effect of the chemical gardens reactions was not explored. Imaging of the self-propulsion of micro-boats with a thermal camera demonstrated that the reaction is exothermic. When PVC tubing containing both $FeCl_3$ seed crystals and camphor was placed on the surface of water, a marked increase in temperature at the end of the tubing containing the seed crystals was registered, as high as 2ºC, which may also promote Marangoni thermo-capillary flows.[54]

However, the effect of heating relatively quick decayed with time, as shown in Supplementary Fig. S3a. Spatial and temporal distributions of temperature are shown in Supplementary Fig. S3b. It is noteworthy that this effect of heating does not give rise to the observable thermo-capillary Marangoni flow preventing self-propulsion.

**Conclusions**

Surface-tension driven of micro-swimmers is important in view of numerous medical and engineering (microfluidics and sensing) applications.[3-9,55-57] We report the use of a self-propelled camphor micro-boat for carrying out a chemical garden reaction. Polymer tubing was filled with camphor at one end and with $FeCl_3$ seed crystals at the other end. The system was placed carefully on the surface of the aqueous solution of potassium hexacyanoferrate ($K_4Fe(CN)_6$). The tubing was propelled by the "camphor engine", and concurrently the chemical garden reaction took place. The reaction gave rise to the "inverse chemical gardens" growing from the top downward, similar to those reported in Ref. 44, and resulted in the formation of heliciform precipitates. The evolution of the gardens was slowed with an increase in the concentration of potassium hexacyanoferrate.

A phenomenological model describing the self-locomotion of the micro-boat is proposed. The model involves two characteristic times, namely the characteristic time scale of the change in the surface tension due to the dissolution of camphor, followed by its diffusion, denoted as $\tau_d$, and the characteristic time of the viscous dissipation $\tau_{fr}$, treated as fitting parameters. Fitting of the models to the experimental data supplied for $\tau_d$ values in the range of $30-200s$, which are close to the time scale of the jump in the surface pressure driving camphor disks, established experimentally by Suematsu *et al*. in Ref. 21. The characteristic time scale of the viscous dissipation slowing the boat was established as $\tau_{fr} \approx 1-4s$ for various solution concentrations and lengths of the micro-boat. This is reasonably explained by the fact that the viscosity and surface tension of the aqueous potassium hexacyanoferrate solutions are only slightly dependent on their concentrations (within the range of studied concentrations, i.e. *c*=0.125-3 wt.%). The exothermic nature of the chemical gardens' reaction is reported. The reported results may serve as a platform for development of chemical reactions guided by self-propelled bodies, driven by surface flows.

**ACKNOWLEDGMENTS**

The authors are indebted to Professor Gene Whyman for fruitful discussions. Acknowledgement is made to the donors of the Israel Ministry of Absorption for the partial support of the scientific activity of Dr. Mark Frenkel.

**AUTHOR CONTRIBUTIONS**

E.B. analyzed data and proposed experimental and theoretical concepts. F.M. planned and carried out
experiments. V.M. planned and carried out experiments, analyzed data. R.G. planned and carried out experiments. A.M. planned and carried out experiments. All authors reviewed the manuscript.

**ADDITIONAL INFORMATION**

**Supplementary information** accompanies this paper at http://www.nature.com/srep
**Competing financial interests**: The authors declare no competing financial interests.